\documentclass[%
  preprint,
 showpacs,
 showkeys,
 preprintnumbers,
 amsmath,amssymb,
 aps,
 prl,
  longbibliography,
 ]{revtex4-1}

\usepackage{amssymb,amsthm,amsmath}
\usepackage[left]{eurosym}

\theoremstyle{definition}

\theoremstyle{remark}

\usepackage{tikz}
\usepackage[breaklinks=true,colorlinks=true,anchorcolor=blue,citecolor=blue,filecolor=blue,menucolor=blue,pagecolor=blue,urlcolor=blue,linkcolor=blue]{hyperref}
\usepackage{graphicx}% Include figure files
\usepackage{url}

\usepackage{xcolor}

\begin{document}

\title{Quantitative easing is an incomplete strategy that must be accompanied by the nullification of debt}

%\cdmtcsauthor{Karl Svozil}
%\cdmtcsaffiliation{Vienna University of Technology}
%\cdmtcstrnumber{407}
%\cdmtcsdate{September 2011}
%\coverpage

\author{Karl Svozil}
\affiliation{Institute for Theoretical Physics, Vienna
    University of Technology, Wiedner Hauptstra\ss e 8-10/136, A-1040
    Vienna, Austria}

\email{svozil@tuwien.ac.at} \homepage[]{http://tph.tuwien.ac.at/~svozil}
%\thanks{ }

%\preprint{CDMTCS preprint nr. 407/2011}

\begin{abstract}
Compound interest as well as inflation grows exponentially with time, whereas other means to repay debt grow polynomially. For this and other, mostly political, reasons, debt without inflation is unsustainable. We suggest a discontinuous way to eliminate debt by nullifying it. This scenario is preferable to current central bank strategies of quantitative easing because it allows the disposal of debt without hyperinflation or bloated balance sheets.
\end{abstract}

\maketitle

\section{A brief analysis of sustainability of debt}

For economies with fractional reserve-generated fiat money \cite{Galbraith-money},
balancing the budget is characterized by
an {\em exponential} growth $D(t) \approx D_0(1+r)^t$ of any initial debt $D_0$ subjected to interest $r$ as a function of time $t$
due to the compound interest; a fact known since antiquity \cite{merton-68}.
At the same time, besides default,
this increasing debt can only be reduced by the following four, mostly {\em linear},  measures:
(i) more income or revenue  $I$ (in the case of sovereign debt: higher taxation or higher tax base);
(ii) less spending $S$;
(iii) increase of borrowing $L$;
(iv) acquisition of external resources $A$ (e.g., through war \cite{smith-gold1}, marriage, or inheritance).
(v) inflation; that is, devaluation of money \cite{mackey-1841,white-1933}.
Whereas (i), (ii) and (iv) without inflation are essentially measures contributing linearly (or polynomially) to the acquisition or compensation of debt,
inflation also grows exponentially with time $t$ at some (supposedly constant) rate $f \ge 1$;
that is, the value of an initial debt $D_0$, without interest ($r=0$),
in terms of the initial values, gets reduced to
$F(t)= D_0 / f^t$.
Conversely, the capacity of an economy to compensate debt will increase with compound inflation:
for instance, the initial income or revenue $I$ will, through adaptions, usually increase exponentially with time in an inflationary regime by $I f^t$.

Because these are the only possibilities, we can consider such economies as closed systems (with respect to money flows),
characterized by the (continuity) equation
\begin{equation}
\begin{split}
I f^t+S+L \approx  D_0(1+r)^t, \textrm{ or} \\
L \approx  D_0\left(1+r\right)^t  - I f^t - S
.
\end{split}
\end{equation}

Let us concentrate on sovereign debt and briefly discuss the fiscal, social and political options.
With regards to the five ways to compensate debt the following assumptions will be made:
First, in non-despotic forms of governments (e.g., representative democracies and constitutional monarchies),
increases of taxation, related to  (i),
as well as spending cuts, related to  (ii),
are very unpopular, and can thus be enforced only in very limited, that is polynomial, forms.

Second, the acquisition of external resources, related to (iv), are often blocked for various obvious reasons;
including military strategy limitations, and lack of opportunities.
We shall therefore disregard the acquisition of external resources entirely and set $A=0$.

As a consequence, without inflation (i.e., for $f=1$),
the increase of debt
\begin{equation}
L \approx  D_0 (1+r) ^t  - I - S
.
\end{equation}
grows exponentially.
This is only ``felt'' after trespassing a quasi-linear region
for which, due to a Taylor expansion around $t=0$,
$D(t) = D_0 (1+r) ^t \approx D_0 + D_0 rt$.

We conclude, that, under the political and social assumptions made, {\em compound debt without inflation is unsustainable.}
Furthermore, inflation, with all its inconvenient consequences and (re-)appropriation,
seems inevitable for the continuous existence of economies based on fractional reserve generated fiat money;
at least in the long run.

\section{Why reserve-generated fiat money rules}

Before proceeding with another, discontinuous solution to issues related to compound interest,
let us shortly discuss and reject at least two ``exotic'' monetary scenarios that have been proposed:
First, consider a commodity-based monetary system, such as one based on gold.
Second, one may maintain fiat money, but consider its creation via interest-free credit,
thereby essentially eliminating interest rates $r=0$ and thus also exponential growth of compound interest, as $1^t=1$.

One of the drawbacks of commodity based money is the limit to liquidity and expansion of the volume of money it can provide
(it could be suspected that rare commodities grow linear or with low polynomiality),
thereby seriously hampering such economies,
and making them less competitive than, and vulnerable to,
other types of monetary schemes allowing  essentially unlimited expansion of credit.

Interest free monetary installments
(which have a long tradition;
cf. the Christian {\it usury}~\citep{noonan-57,Weiss-Zinswucher-Christ}, and the Islamic {\it riba}~\citep{Hanke-Zinswucher-Islam})
 are characterized by ``infinitely cheap'' seed money, yielding to excessive
borrowing and economic bubbles.
As a result, the volume of money must be severely restricted by measures and criteria which are highly questionable,
and mostly will not reflect supply-and-demand-sided market self-regulations.
Also, it can be expected that privilege or a certain arbitrariness and chance will govern the distribution
of such interest-free money; and therefore will result in questionable (re-)distribution of resources.

In contradistinction, it is not totally unreasonable to speculate
that the interest levied in reserve-generated fiat monetary serves as a selection mechanism,
thereby (re-)appropriating economic resources and wealth through market mechanisms.
We therefore suspect that the current reserve-generated fiat money, besides its long-term unsustainability discussed earlier,
is preferable over other forms of money.

\section{Appropriation of wealth}

The current crisis of mayor currencies, in particular the Euro, appears to demonstrate that,
in order to be intrinsically political, social and economic sustainable,
every currency zone formed by societies using one and the same money must also have the same economic, social and economic framework,
including equal taxation, and the capacity and will to enforce taxation,
as well as the same level of corruption.
(We assume that it is not totally unreasonable to suspect that corruption cannot be avoided,
so it has to be accepted at some levels.)

For instance, China, Germany or Austria seem to be a ``rich countries with relative poor people,''
whereas Spain or Italy or France appear to be ``poor countries with rich people.''
This may be due to a lot of factors, including the sorts and percentage of taxation or the retirement system.
Alas the resulting relative imbalances in sovereign debt may build up to huge social tensions
when it comes to ``rescue packages'' effectively transferring wealth from those underprivileged who are heavily taxed to the privileged with
low taxation.
This may cause tensions in both ways, because the underprivileged do not want to increase the wealth of the privileged;
and the privileged do not want to acknowledge their privilege, and
therefore demand to be rescued by the underprileged.

In Europe, mmbalances are further increased ``accidentally'' by an ill-conceived  interbank payment
and clearance system for the real-time processing of cross-border transfers
(TARGET2) in the Euro-zone, which essentially presents the possibility of unlimited automatic increase of inter-state loans and debts
inside of the Euro-zone. Unlike in the US Federal reserve system, this is not balanced regularly by the exchange of collaterals,
but allowed to grow unlimited. As a result, debt levels of quasi-insolvent states (with little access to private borrowing) has increased dramatically,
with little chance for the debtors (mainly the German Bundesbank) to recover those quasi-loans
(which, by the way, perform at very low interest rates)
through inverse transactions \cite{Sinn-Target2}.

These factors -- widely varying levels of taxation, social spending and corruption, as well as an ill-conceived TARGET2-system --
have contributed to a sort of ``cargo culture'' in the ``debtor nations,''
which try to maintain and defend their consumption and spending levels ``at all costs;''
mostly through ceremonies such as the public demonstration of anger, and through neglect.
At the same time, the ``debtor nations'' demand to be rescued by their ``creditors.''
The  ``creditors'' become increasingly frustrated and
unwilling to support this ``cargo culture'' through unsustainable levels of debt and the TARGET2 system,
combined with high corruption and small tax rates of debtor nations.
This is met with aggressive popular outrage by their debtors.
Both frustrations of debtors and creditors contribute to potentially dangerous political tensions among nations.

As we will be dealing mainly with a proposal for a solution to the debt crises, we just mention that,
despite the urgent necessity to resolve the deficiencies of the Eurosystem TARGE2 scheme,
for any common monetary system to be sustainable a necessary criterion is the {\em harmonization and conformity
of taxation, social as well as economic usances, including levels of corruption and appropriation of wealth.}

\section{Nullification strategy}

In what follows we shall propose a discontinuous, and relatively ``benign'' way to reduce debt levels without continuous inflation.
Historically, unsustainable levels of debt have lead to business cycles, which some believe are unavoidable and inevitable \cite{ReinhartRogoff}.
In contradistinction to these rather gloomy monetary frameworks, we will propose a very simple solution to the debt crises,
eliminating it by basically collecting sufficient portions of credit certificates into a ``bad bank'' -- which could be a portion of
a central bank or any other institution capable of fiat money creation -- followed by a formal nullification (``haircut to zero'') of this debt.

The proposed scheme involves three phases:
(i) In the first, foundational phase, a ``bad bank'' (BB) is formally created for the sole task to absorb ``bad loans'' and ``unsustainable debt levels.''
This institution must have the capacity to create fiat money in return for credit certificates, and not subjected to conventional rating proceduress.
It might be preferable to associate or adjoin such a bad bank with some international monetary institution,
such as the {\it International Monetary Fund} (IMF),
or the {\it Bank for International Settlements} (BIS).

(ii) In the second, acquisition phase, BB acquires credit certificates in return for newly created fiat money by negotiating with markets;
if possible by realizing a ``haircut,''
thereby making it possible for endangered creditors (such as banks or funds) to get rid of their ``bad loans.''
The duration period of this phase is announced well in advanced,
to allow all such settlements to take place.

(iii) In the third, closing phase, BB is liquidated. Thereby the debt is either reduced to sustainable levels or entirely nullified.

Let us discuss some arguments against such ``haircuts.''

First, the scheme encourages unbounded creation of debt, as it amounts to the message:
``the more debt is created in return for consumption, the higher is the (re-)appropriation of wealth toward the debtor.''
This argument appears to be correct.
Alas with respect to (re-)appropriation the situation is limited by the regulations that can be imposed;
and throughout history there have already been very similar trade imbalances and inequalities building up.
Very seldom these trade imbalances could be settled without war or nullification of debt.
So, even the current regimes do not appear to resolve these issues.

Second, one may favor an apocalyptic (black Friday type) ``crash scenario''
motivated by the clearing of suboptimal economic structures at the end of a business cycle.
In such a meltdown, all imbalances and debt-creditors are whipped out by a general liquidation of all moneys, followed by a restructuring of the economy.
Any such ``Armageddon'' appears to be extremely risky and dangerous to the well-being of individuals and nations,
as it bears the potential for great political and social unrest and tragedy.

Third, it may not be imprudent to speculate that the liquidation of BB will not result in overall inflation  \cite{anderson-2010-doubling} --
although certain segments of an economy might suffer from it --
as inflation and higher price fixes in markets occur through
scarcity of goods combined with a non-scarcity of money, resulting in subjective fears of price increases on the consumer's side.
At present, there is no scarcity of goods in sight; even in the presence of diminishing energy and other resources.
Indeed, regularly nullifying debt in a controlled way presents a good alternative to high inflation.

Fourth, imagine the benefit and potentiality of all nations starting with zero debt levels again;
as compared to having to fight the unfolding exponential expansion of compound interest, a war that is provable
impossible to win.

\section{Comparison with strategy of ``quantitative easing''}

Finally, one may argue that this is a direct realization of the printing press;
thereby allowing the ``government and other debtors to acquire unlimited amounts of money.''
Indeed, any government, and politicians in general, might be tempted to go on an unlimited spending spree without levying
any taxes at all.
This argument is essentially correct.
In order to avoid such behavior, restrictions to government spending and (easing) taxation have to be installed.
Alas, as has been argued earlier, for political, social and psychological reasons,
even in the current monetary regimes embedded in politics,
any such attempts are preliminary at best and will eventually fail.

Indeed, a very similar but ``incomplete'' approach -- effectively realizing steps (i) and (ii) but not (iii),
-- termed ``quantitative easing'' (QE) -- is pursued by virtually all central banks.
For instance, on April 4th, 2013, the Bank of Japan unleashed \cite{BoJ-2013-qe}
``a new phase of monetary easing both in terms of quantity and quality.''
In the United States of America a former Chairman of the Federal Reserve declared that ``we can always print [[more]] money.''
And, to present another anecdote, the Eurosystem has just granted Ireland relieve in pushing repayments of about 40 billion \EUR
to an average maturity of over 34 years.

In that way, those central banks have already become BBs ``for all practical purposes.''
Thereby, governments might hope to obtain compensation for the exponentially rising levels of sovereign debt
by payments of the (compound) interest the central banks receive from the very bonds the government issued.
Although such schemes are principally unbounded, to some, say rating agencies or investors,
the necessary exponentially (fiat) money volumes created
may eventually appear unsustainable at certain levels.
Indeed it may eventually dawn upon everybody
that repayment of this debt in the exponential compound interest regime
will never be possible by economic growth, and is illusory without hyperinflation or default.

We thus observe that this incomplete nullification of (sovereign)
debt renders economies increasingly volatile and vulnerable to the follies of the market.
Because even with unbounded QE without nullification,
there is a huge danger of carrying through exponential schemes of compound interest, and thus of debt.

%\begin{acknowledgments}
%\end{acknowledgments}

%\bibliography{svozil}

\begin{thebibliography}{12}%
\makeatletter
\providecommand \@ifxundefined [1]{%
 \@ifx{#1\undefined}
}%
\providecommand \@ifnum [1]{%
 \ifnum #1\expandafter \@firstoftwo
 \else \expandafter \@secondoftwo
 \fi
}%
\providecommand \@ifx [1]{%
 \ifx #1\expandafter \@firstoftwo
 \else \expandafter \@secondoftwo
 \fi
}%
\providecommand \natexlab [1]{#1}%
\providecommand \enquote  [1]{``#1''}%
\providecommand \bibnamefont  [1]{#1}%
\providecommand \bibfnamefont [1]{#1}%
\providecommand \citenamefont [1]{#1}%
\providecommand \href@noop [0]{\@secondoftwo}%
\providecommand \href [0]{\begingroup \@sanitize@url \@href}%
\providecommand \@href[1]{\@@startlink{#1}\@@href}%
\providecommand \@@href[1]{\endgroup#1\@@endlink}%
\providecommand \@sanitize@url [0]{\catcode `\\12\catcode `\$12\catcode
  `\&12\catcode `\#12\catcode `\^12\catcode `\_12\catcode `\%12\relax}%
\providecommand \@@startlink[1]{}%
\providecommand \@@endlink[0]{}%
\providecommand \url  [0]{\begingroup\@sanitize@url \@url }%
\providecommand \@url [1]{\endgroup\@href {#1}{\urlprefix }}%
\providecommand \urlprefix  [0]{URL }%
\providecommand \Eprint [0]{\href }%
\providecommand \doibase [0]{http://dx.doi.org/}%
\providecommand \selectlanguage [0]{\@gobble}%
\providecommand \bibinfo  [0]{\@secondoftwo}%
\providecommand \bibfield  [0]{\@secondoftwo}%
\providecommand \translation [1]{[#1]}%
\providecommand \BibitemOpen [0]{}%
\providecommand \bibitemStop [0]{}%
\providecommand \bibitemNoStop [0]{.\EOS\space}%
\providecommand \EOS [0]{\spacefactor3000\relax}%
\providecommand \BibitemShut  [1]{\csname bibitem#1\endcsname}%
\let\auto@bib@innerbib\@empty
%</preamble>
\bibitem [{\citenamefont {Galbraith}(1975)}]{Galbraith-money}%
  \BibitemOpen
  \bibfield  {author} {\bibinfo {author} {\bibfnamefont {John~Kenneth}\
  \bibnamefont {Galbraith}},\ }\href@noop {} {\emph {\bibinfo {title} {Money.
  {W}hence It Came, Where It Went}}}\ (\bibinfo  {publisher} {Andr\`e Deutsch
  Limited},\ \bibinfo {address} {London},\ \bibinfo {year} {1975})\BibitemShut
  {NoStop}%
\bibitem [{\citenamefont {Merton}(1968)}]{merton-68}%
  \BibitemOpen
  \bibfield  {author} {\bibinfo {author} {\bibfnamefont {Robert~K.}\
  \bibnamefont {Merton}},\ }\bibfield  {title} {\enquote {\bibinfo {title} {The
  {M}atthew effect in science},}\ }\href {\doibase 10.1126/science.159.3810.56}
  {\bibfield  {journal} {\bibinfo  {journal} {Science}\ }\textbf {\bibinfo
  {volume} {159}},\ \bibinfo {pages} {56--63} (\bibinfo {year}
  {1968})}\BibitemShut {NoStop}%
\bibitem [{\citenamefont {{Smith Jr.}}(1989)}]{smith-gold1}%
  \BibitemOpen
  \bibfield  {author} {\bibinfo {author} {\bibfnamefont {Arthur~Lee}\
  \bibnamefont {{Smith Jr.}}},\ }\href@noop {} {\emph {\bibinfo {title}
  {{H}itler's Gold. {T}he Story of the {N}azi War Loot}}}\ (\bibinfo
  {publisher} {Berg},\ \bibinfo {address} {Oxford, New York, Munich},\ \bibinfo
  {year} {1989})\BibitemShut {NoStop}%
\bibitem [{\citenamefont {Mackay}(1841)}]{mackey-1841}%
  \BibitemOpen
  \bibfield  {author} {\bibinfo {author} {\bibfnamefont {Charles}\ \bibnamefont
  {Mackay}},\ }\href {http://www.gutenberg.org/files/24518/24518-h/dvi.html}
  {\emph {\bibinfo {title} {Memoirs of Extraordinary Popular Delusions and the
  Madness of Crowds. {V}olume {I}}}}\ (\bibinfo  {publisher} {Office of the
  National Illustrated Library},\ \bibinfo {address} {London},\ \bibinfo {year}
  {1841})\BibitemShut {NoStop}%
\bibitem [{\citenamefont {White}(1933)}]{white-1933}%
  \BibitemOpen
  \bibfield  {author} {\bibinfo {author} {\bibfnamefont {Anrew~Dickson}\
  \bibnamefont {White}},\ }\href
  {http://http://mises.org/books/inflationinfrance.pdf} {\emph {\bibinfo
  {title} {Fiat Money Inflation in {F}rance. {H}ow it came, what it brought,
  and how it ended}}}\ (\bibinfo  {publisher} {D. Appelton-Century Company},\
  \bibinfo {address} {New York, London},\ \bibinfo {year} {1933})\BibitemShut
  {NoStop}%
\bibitem [{\citenamefont {{Noonan, Jr.}}(1957)}]{noonan-57}%
  \BibitemOpen
  \bibfield  {author} {\bibinfo {author} {\bibfnamefont {John~Thomas}\
  \bibnamefont {{Noonan, Jr.}}},\ }\href@noop {} {\emph {\bibinfo {title} {The
  scholastic analysis of usury}}}\ (\bibinfo  {publisher} {Harvard University
  Press},\ \bibinfo {address} {Cambridge, Mass.},\ \bibinfo {year}
  {1957})\BibitemShut {NoStop}%
\bibitem [{\citenamefont {Wei{\ss}}(2005)}]{Weiss-Zinswucher-Christ}%
  \BibitemOpen
  \bibfield  {author} {\bibinfo {author} {\bibfnamefont {Andreas~M.}\
  \bibnamefont {Wei{\ss}}},\ }\bibfield  {title} {\enquote {\bibinfo {title}
  {{Z}insen und {W}ucher. {D}as kirchliche {Z}insverbot und die {H}indernisse
  auf dem {W}eg zu seiner {K}orrektur},}\ }in\ \href {\doibase
  10.1007/3-211-28108-8_5} {\emph {\bibinfo {booktitle} {{G}eld- und
  {K}reditwesen im {S}piegel der {W}issenschaft}}},\ \bibinfo {editor} {edited
  by\ \bibinfo {editor} {\bibfnamefont {Ulrike}\ \bibnamefont {Aichhorn}}}\
  (\bibinfo  {publisher} {Springer},\ \bibinfo {address} {Vienna},\ \bibinfo
  {year} {2005})\ pp.\ \bibinfo {pages} {123--156}\BibitemShut {NoStop}%
\bibitem [{\citenamefont {Hanke}(2005)}]{Hanke-Zinswucher-Islam}%
  \BibitemOpen
  \bibfield  {author} {\bibinfo {author} {\bibfnamefont {Marcus}\ \bibnamefont
  {Hanke}},\ }\bibfield  {title} {\enquote {\bibinfo {title} {{Z}insverbot und
  islamische {B}ank. {V}on {D}atteln und {K}reditkarten},}\ }in\ \href
  {\doibase 10.1007/3-211-28108-8_6} {\emph {\bibinfo {booktitle} {{G}eld- und
  {K}reditwesen im {S}piegel der {W}issenschaft}}},\ \bibinfo {editor} {edited
  by\ \bibinfo {editor} {\bibfnamefont {Ulrike}\ \bibnamefont {Aichhorn}}}\
  (\bibinfo  {publisher} {Springer},\ \bibinfo {address} {Vienna},\ \bibinfo
  {year} {2005})\ pp.\ \bibinfo {pages} {157--175}\BibitemShut {NoStop}%
\bibitem [{\citenamefont {Sinn}\ and\ \citenamefont
  {Wollmersh\"auser}(2012)}]{Sinn-Target2}%
  \BibitemOpen
  \bibfield  {author} {\bibinfo {author} {\bibfnamefont {Hans-Werner}\
  \bibnamefont {Sinn}}\ and\ \bibinfo {author} {\bibfnamefont {Timo}\
  \bibnamefont {Wollmersh\"auser}},\ }\bibfield  {title} {\enquote {\bibinfo
  {title} {Target loans, current account balances and capital flows: the
  {ECB}'s rescue facility},}\ }\href {\doibase 10.1007/s10797-012-9236-x}
  {\bibfield  {journal} {\bibinfo  {journal} {International Tax and Public
  Finance}\ }\textbf {\bibinfo {volume} {19}},\ \bibinfo {pages} {468--508}
  (\bibinfo {year} {2012})}\BibitemShut {NoStop}%
\bibitem [{\citenamefont {Reinhart}\ and\ \citenamefont
  {Rogoff}(2009)}]{ReinhartRogoff}%
  \BibitemOpen
  \bibfield  {author} {\bibinfo {author} {\bibfnamefont {Carmen~M.}\
  \bibnamefont {Reinhart}}\ and\ \bibinfo {author} {\bibfnamefont {Kenneth~S.}\
  \bibnamefont {Rogoff}},\ }\href@noop {} {\emph {\bibinfo {title} {This Time
  Is Different: Eight Centuries of Financial Folly}}}\ (\bibinfo  {publisher}
  {Princeton University Press},\ \bibinfo {address} {Princeton, NJ},\ \bibinfo
  {year} {2009})\BibitemShut {NoStop}%
\bibitem [{\citenamefont {Anderson}\ \emph {et~al.}(2010)\citenamefont
  {Anderson}, \citenamefont {Gascon},\ and\ \citenamefont
  {Liu}}]{anderson-2010-doubling}%
  \BibitemOpen
  \bibfield  {author} {\bibinfo {author} {\bibfnamefont {Richard~G.}\
  \bibnamefont {Anderson}}, \bibinfo {author} {\bibfnamefont {Charles~S.}\
  \bibnamefont {Gascon}}, \ and\ \bibinfo {author} {\bibfnamefont {Yang}\
  \bibnamefont {Liu}},\ }\bibfield  {title} {\enquote {\bibinfo {title}
  {Doubling your monetary base and surviving: Some international experience},}\
  }\href {http://research.stlouisfed.org/publications/review/article/8509}
  {\bibfield  {journal} {\bibinfo  {journal} {Federal Reserve Bank of St. Louis
  Review}\ }\textbf {\bibinfo {volume} {92}},\ \bibinfo {pages} {481--506}
  (\bibinfo {year} {2010})}\BibitemShut {NoStop}%
\bibitem [{\citenamefont {of~Japan}(April 4, 2013)}]{BoJ-2013-qe}%
  \BibitemOpen
  \bibfield  {author} {\bibinfo {author} {\bibfnamefont {Bank}\ \bibnamefont
  {of~Japan}},\ }\href
  {http://www.boj.or.jp/en/announcements/release\_2013/k130404a.pdf} {\enquote
  {\bibinfo {title} {Introduction of the ``quantitative and qualitative
  monetary easing''},}\ } (\bibinfo {year} {April 4, 2013}),\ \bibinfo {note}
  {statements on Monetary Policy 2013}\BibitemShut {NoStop}%
\end{thebibliography}


\end{document}